\documentclass[aps,pre,preprint,groupedaddress]{revtex4}
\usepackage{graphicx,amssymb,amsmath}
\makeatother      
\newlength{\newwidth}
 
\begin{document}

\preprint{}

\title{Phase Transition in the ABC model}


\author{M. Clincy$^1$, B. Derrida$^2$, M. R. Evans$^1$}
\affiliation{$^1$School of Physics, University of Edinburgh,
Mayfield Road, Edinburgh EH9 3JZ, UK\\$^2$Laboratoire de Physique Statistique
de l'Ecole Normale Sup\'erieure, 24 Rue Lhomond, 75231 Paris Cedex 05, France}


\date{\today}

\begin{abstract}
Recent studies have shown that one-dimensional driven systems can
exhibit phase separation even if the dynamics is governed by local
rules.
The ABC model, which comprises three particle species that
diffuse asymmetrically around a ring, shows anomalous coarsening into
a phase separated steady state. 
In the limiting
case in which the dynamics is symmetric and the parameter $q$
describing the asymmetry tends to one, no phase separation occurs and
the steady state of the system is disordered. In the present work we
consider the weak asymmetry 
regime $q=\exp{(-\beta/N)}$  where $N$ is the system size and study how the
disordered state is approached.
In the case of equal densities, we find 
that the system exhibits a second order phase
transition   at some nonzero $\beta_c$.
 The value of $\beta_c = 2 \pi \sqrt{3}$ and the optimal
profiles  can be obtained   by
writing the 
exact large
deviation functional.
For nonequal densities, we write down mean field equations and analyze
some of their predictions.
\end{abstract}


			      \maketitle

\newcommand{\del}{\partial}
\newcommand{\ul}{\underline}
\newcommand{\im}{{\rm i}}
\newcommand{\be}{\begin{eqnarray}}
\newcommand{\ee}{\end{eqnarray}}
\newcommand{\ba}{\begin{array}}
\newcommand{\ea}{\end{array}}
\newcommand{\f}[2]{\frac{#1}{#2}}
\newcommand{\mc}{\mathcal}
\newcommand{\bs}{\backslash}
\newcommand{\bra}[1]{\langle{#1}\!\mid}
\newcommand{\ket}[1]{\mid\!{#1}\rangle}
\newcommand{\D}{\ensuremath{\mathrm{d}}}


\section{Introduction}

Nonequilibrium steady states, wherein the properties of a system are
stationary but the steady state probabilities are not described by
Boltzmann weights with respect to a local energy function, may exhibit
a number of interesting phenomena absent from equilibrium systems:
for example,
driven diffusive systems \cite{KLS}
have generically long-range correlations;
phase
transitions may also occur in one-dimensional non-equilibrium steady states
although they  are of course precluded from
one-dimensional equilibrium systems with short-range interactions.

Examples of nonequilibrium phase transitions include the absorbing
state phase transitions \cite{Haye} and boundary induced phase
transitions in driven systems wherein a conserved current is driven
through a finite open system \cite{krug,DDM,DEHP,SD,EFGM}.
Bulk (i.e. not boundary-driven) phase
transitions may arise in systems with no absorbing states
like conserving driven systems through the
introduction of several species of particles
\cite{DJLS,Mallick,Evans96,AHR,KLMT,EKLM}.

Phase separation 
has been exhibited in several
one dimensional driven
systems 
\cite{EKKM,LBR,AHR}
consisting of
several species of  particles with nearest
neighbour exchanges occurring with prescribed rates.
Models in this class are
the ABC model  \cite{EKKM}, the AHR model \cite{AHR,RSS,KLMT}
-- which both contain three species of
particles -- and the LR model 
\cite{LBR} which consists of two sublattices with two species
each. The phase separation in these models has the striking feature of
the domains of each species being pure.  That is, far away from the
domain walls, there is zero probability of finding a particle of one
species in a domain of a different species. This is referred to as
strong phase separation.

An understanding of this phenomenon has  emerged through the exact
solution of the steady state in some special cases of the
ABC\cite{EKKM} and the LR model\cite{LBR}.  Even though the dynamics
is strictly local, in these special cases it has been shown that the
steady state obeys detailed balance with respect to a long-range
energy function.  The long-range interaction leads to superextensivity
of this energy function i.e. the energy of most microscopic
configurations scales quadratically with system size $N$
so that the contribution of the entropy becomes negligible.
Almost all
configurations are therefore suppressed   and only
strongly phase separated configurations contribute.  Although
generally in the ABC and the LR models  for non-equal numbers of
particles  detailed balance does not hold and
one does not have an energy function, the same strong phase separation
is  observed  in simulations \cite{EKKM}.

In the ABC model there is a parameter $q$ that governs the local
dynamics. It describes the asymmetry in the 
rates of nearest neighbour particle exchanges 
(see section~\ref{ABCdef}). If $q < 1$ is held fixed then
one always has strong phase separation
in the thermodynamic limit ($N\to\infty$).
  However for $q=1$, in which case the particles exchange 
symmetrically, the system is in a homogeneous
disordered state where all configurations are equally probable.  In
the present work we investigate the ABC model in what we shall refer
to as the 
weak asymmetry
regime.  That is, we introduce a system size
dependence into $q$ so that an extensive energy is recovered (when
an  energy function indeed exists).  It turns out
that the appropriate choice of $q$ is 
\be 
q = \exp{\left(-\f{\beta}{N}\right)}\; .
\label{scale1}
\ee 
Thus $\beta$ is now the control parameter and plays the role of an
inverse temperature i.e. $\beta = 1/T$.  In this 
 regime an
interesting question is as to how the transition from the strongly
phase separated state to the disordered state occurs as $T$ is varied.
Since the energy function is long range it is possible to have a phase
transition  at a well defined
temperature $T_c =  \beta_c^{-1}$ even though the system is one dimensional
(phase transitions in fact do occur in one-dimensional systems with
algebraically decaying interaction \cite{FMN} or in the Bernasconi
model \cite{Bernasconi} which is an Ising spin model
with energy function given by long-range, four spin interactions
\cite{MPR}).

In the present paper, we first summarize in section II some known facts about the ABC
model. In section III, we present our  results 
of Monte Carlo simulations done in 
the weak asymmetry regime  
  for an equal number of particles of each
species.  These simulations 
indicate that there is a critical value $\beta_c = {1 \over T_c} \simeq
11$. 
 In Section IV
 we derive   the exact  free energy functional for an arbitrary density
profile
 from the  expression for the  weights in the steady
state in the case of equal particle numbers.
  By minimizing this functional we obtain  the exact value
of $\beta_c= 2 \pi \sqrt{3} \simeq 10.88\dots$ and the shape of the density
profiles in the neighbourhood of the transition.  
In section V, we investigate the case of  arbitrary  global densities within a
mean field approximation.
We observe that this approximation  turns out  to be exact in the case of
equal numbers of particles.

\section{The ABC model}
\label{ABCdef}
The ABC model is defined on a 1d ring with $N$ lattice sites.  Each
site is occupied by one of three types of particles denoted as $A$,
$B$ or $C$. They exhibit hard-core interaction, i.e. only one particle
per site is allowed.   
Neighboring sites on the ring
are exchanged  according  to the
following rates: 
\be AB & {{q \atop \longrightarrow}\atop
{\longleftarrow \atop 1}} & BA\\ BC & {{q \atop \longrightarrow}\atop
{\longleftarrow \atop 1}} & CB\\ CA & {{q \atop \longrightarrow}\atop
{\longleftarrow \atop 1}} & AC
\label{update}
\ee
Thus for $q\ne 1$ the particles diffuse asymmetrically around the ring 
and in the case $q=1$ they diffuse symmetrically.
 Note that periodic boundary conditions are implied
and the dynamics conserves the number of particles.

This model has been extensively studied in \cite{EKKM} by analytical
and numerical means. 
Let us  summarise  in this section the main results.
Starting from an initially random configuration, the system 
coarsens into a strongly phase separated state for any $q\ne 1$. 
Thus the steady state 
 exhibits long-range order,
even though the dynamics is strictly local. We
will restrict our discussion of $q \ne 1$ to the case $q<1$ for
simplicity. (The case $q>1$ can be obtained by the transformation $q
\to 1/q$ together with the exchange of $A$ and $B$ particles).  For
$q=1$ particles of all species have equivalent dynamics which results
in a steady state in which all configurations have equal weight  and the
steady state of the system is disordered.

To understand the coarsening dynamics for $q < 1$, note that domain walls of the
type $BA$, $CB$ and $AC$ are unstable to interchanges leading to $AB$,
$BC$ and $CA$ respectively.  Therefore
 $A$ particles are driven to
the left in a $B$ domain and to the right in a $C$  domain (particles $B$ and
$C$  are also driven in other species domains).
Thus the system arrives
at a metastable configuration of the form $A\dots AB\dots BC \dots CA
\dots AB\dots BC \dots CA\dots$ and a slow coarsening
process, involving the elimination of the smallest domains, ensues. For
example, the time it takes an $A$ particle to traverse a $B$ domain of
length $l$ is of order $q^l$. Thus the elimination of domains of size
$l$ occurs at a rate of order $q^{-l}$ which results in the typical
domain size growing as $l\sim \ln t$. This growth law which is slower
than any power of $t$ is referred to as anomalous
coarsening\cite{Evans}.  Ultimately the coarsening process results in
a strongly phase separated state comprising three pure domains.

The general steady state is not, as yet, known.
In the special case of equal particle numbers $N_A = N_B = N_C = N/3$,
however,
one can show that the steady state obeys detailed balance with respect
to a long-range 
energy function $\mathcal{H}$. A configuration
of the system is specified
by the set of indicator variables $\{X_i\}=
\{A_i, B_i, C_i\}$ which take values 1 or 0 according to whether
site $i$ is occupied by the relevant particle.
For example $A_i=1$ if site $i$ is occupied by an
$A$ particle.
Clearly, these variables satisfy $A_i+ B_i+ C_i =1$. With these indicator variables
the energy function
may be written as: 
\be 
\mathcal{H}(\{X_i\}) =
 {1 \over N} \sum_{i=1}^N\sum_{k=1}^{N-1} k
\left(B_iC_{i+k}+C_iA_{i+k}+A_iB_{i+k}\right) 
\label{hamiltonian}
\ee
and the steady state weights of the system are given by  
\be
P(\{X_i\}) = Z_N^{-1}q^{\mathcal{H}(\{X_i\})}
\label{weights}
\ee
where $Z_N = \sum q^{\mathcal{H}(\{X_i\})}$ denotes the partition function.
Note that the energy given in \cite{EKKM} differs by a constant
from (\ref{hamiltonian}).

In (\ref{hamiltonian})  the interaction between sites $i$ 
and $i+k$ are both long-range and
asymmetric
and the energy
function is superextensive and scales quadratically with system size
$N$.

The width of the domain walls in the phase separated state is
of order $1/|\ln{q}|$ \cite{EKKM}. So for $q \to 1$ the size of the
domain walls diverges and the system will be in a homogeneous disordered
state. Moreover we expect an interesting  regime to occur
when the width of the domain walls is of the order of the domain lengths
$N/3$ i.e.
\be
\frac{1}{|\ln{q}|} \sim O(N)
\ee
This yields the weakly asymmetric regime stated in the introduction
(\ref{scale1}) where the control
parameter is now $T=1/\beta$.
The steady state weight
(\ref{hamiltonian},\ref{weights}) also
confirms that (\ref{scale1}) is the natural choice of scaling
variable since under this scaling (\ref{weights}) becomes
\be
P_N(\{X_i\}) = Z_N^{-1}\exp{\left(-\beta  E_N(\{X_i\})\right)}
\label{weights2}
\ee
where $Z_N$ is a normalization constant (the partition function)
and 
the extensive rescaled  energy 
$E_N(\{X_i\})$
is defined as
\be
E_N(\{X_i\}) = \mathcal{H}(\{X_i\})/N \; \quad .\label{def-energy}
\ee

\section{Monte Carlo Simulations}
We have measured by Monte-Carlo simulations
 the number of nearest
neighbour pairs of sites occupied by the same species of particles in the
steady state. We
shall refer to these as nearest neighbour (nn) matching pairs.  For a
completely disordered system  (i.e. for $\beta=0$) the probability of finding a nn matching
pair is $1/3$ for large $N$. 
As $\beta$ increases, one expects this number to increase and to be equal
to 1
as $\beta \to \infty$ (i.e. as one reaches the strongly separated
regime).

In figure 1, we show the results of our simulations for $m_N$ defined as
\begin{eqnarray}
m_N &=& \frac{\mbox{number of nn matching pairs}}{\mbox{system
size}} - \f13
\\
&=& \f1N \sum_{i=1}^N\left[\;A_iA_{i+1}
+B_iB_{i+1}+C_iC_{i+1}\right] -\frac{1}{3}\;,
\label{def-order-parameter}
\end{eqnarray}
with
the occupation variable $X_i = A_i, B_i, C_i$ defined as in
(\ref{hamiltonian}).
Note that because of the periodic boundary conditions, site $N+1$
is identified with site $1$.
We see that as $N$ increases, $m_N$ seems to be closer and closer to zero
as
$\beta <  11$ 	whereas it seems to have a well-defined limit $m$  which
depends on $\beta$ for $\beta > 11$
\begin{eqnarray}
m = \lim_{N\to\infty}m_N\;.
\label{def-op-limit}
\end{eqnarray}

Note that $m$ only contains nearest neighbour correlations as opposed
to e.g. the measure of order used in \cite{EKKM} which contains long
range terms.  Also one could consider the lowest Fourier mode of the
density profile (see section IV).

\vspace*{0.5cm}
\begin{figure}[h]
\includegraphics[scale=0.3]{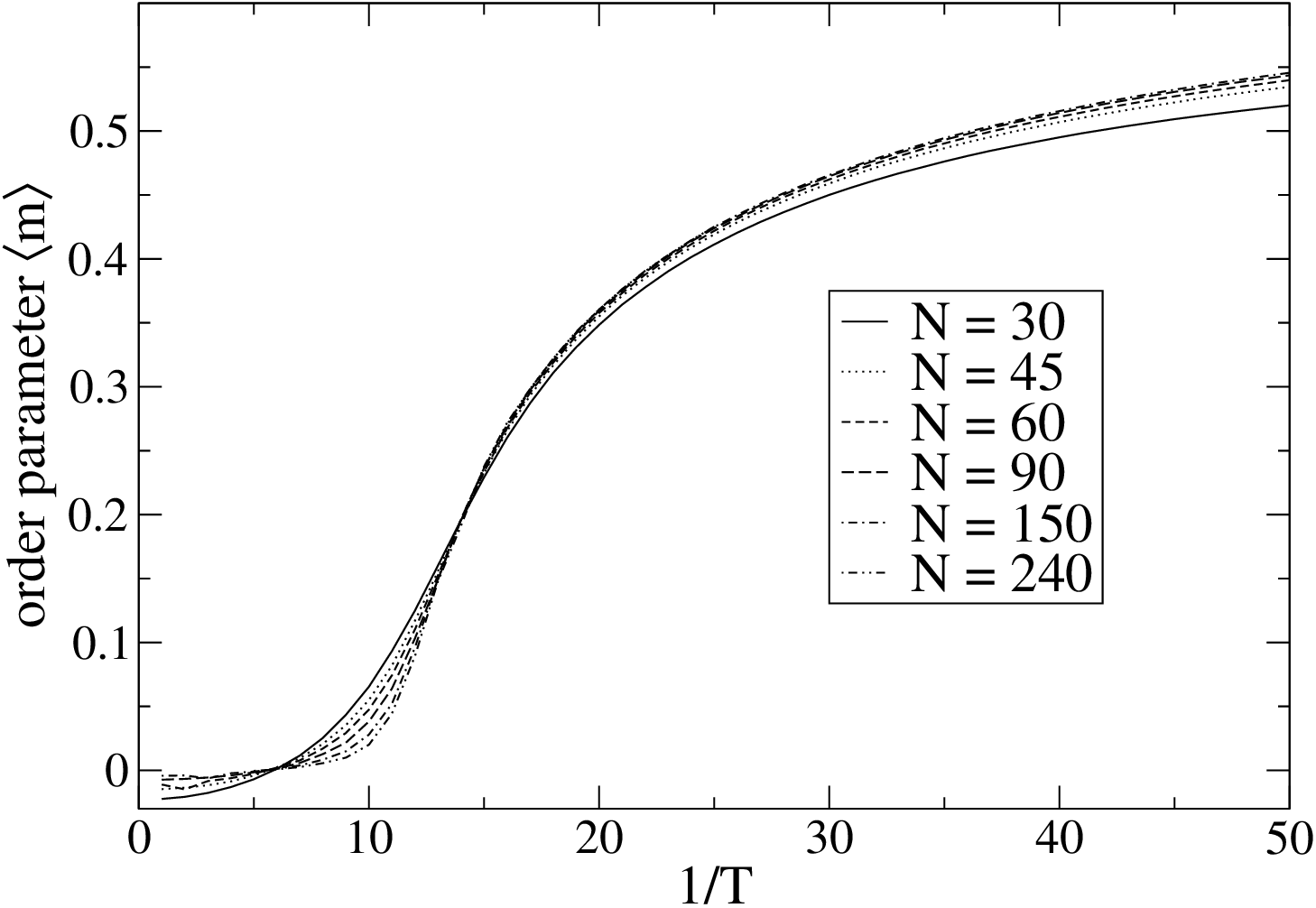}
\caption{Parameter $m$ defined by (\ref{def-order-parameter}) for system
sizes $N= 30$ to $240$\label{nn-pairs}}
\end{figure}
\begin{figure}[h]
\includegraphics[scale=0.3]{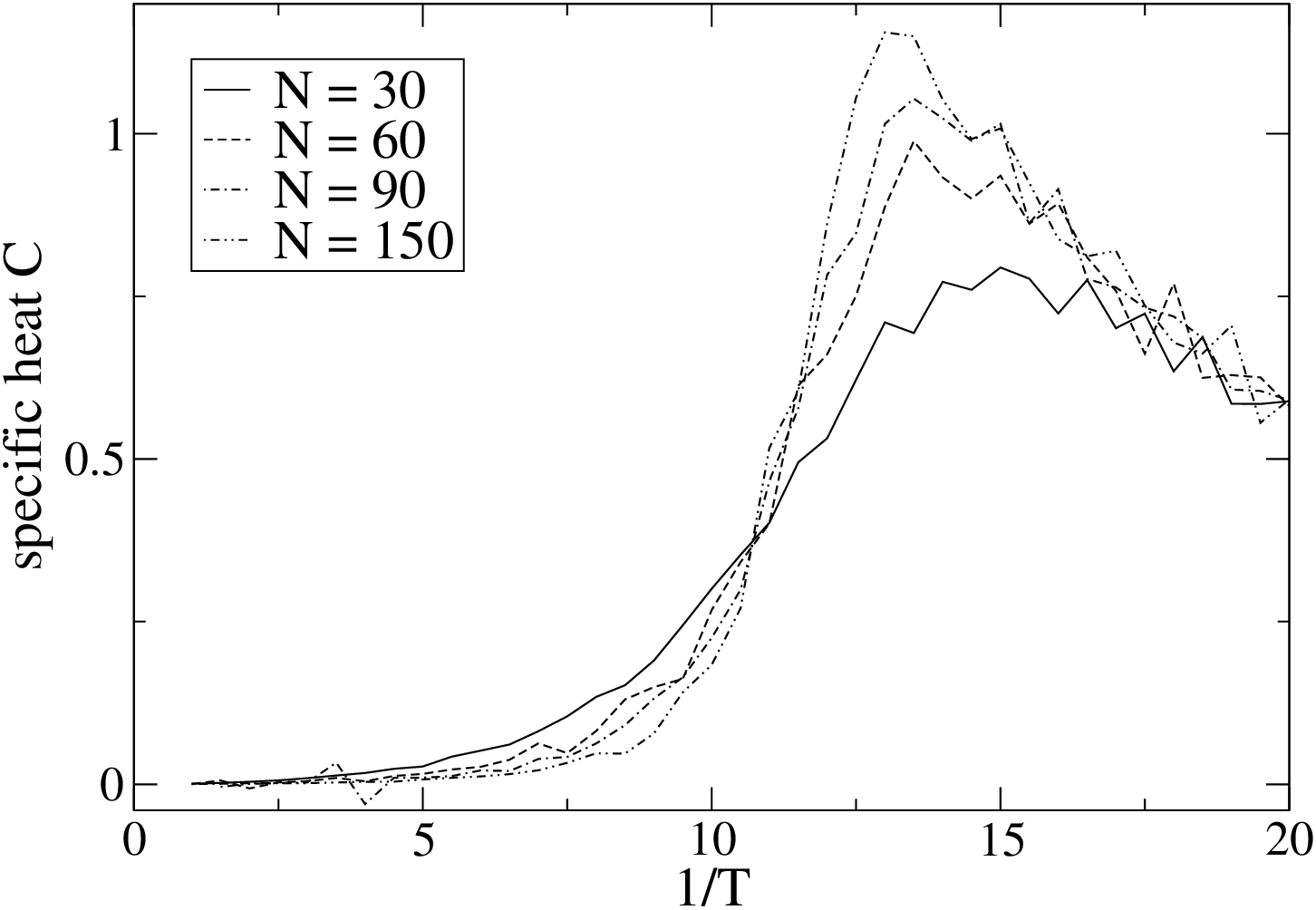}
\caption{Specific heat for system sizes $N= 30,60,90,150$\label{specheat}}
\end{figure}

For $T<T_c$ we expect $0 < m < 2/3$ with $m$
approaching $2/3$ in the limit $T\to 0$.  This behaviour can be seen
in Fig. \ref{nn-pairs}.
For $1/T \approx 11$ a crossover from an ordered to a disordered state
appears which becomes sharper with increasing system size.
 In
Fig. \ref{specheat} we plot the specific heat defined as $C_N = \del
\overline{E_N}/\del T$. One sees strong finite size effects at $\beta=1/T  \approx
11$. Moreover, the curves suggest that a discontinuity emerges in the
infinite system limit which would be consistent with a second-order phase
transition.
\vspace*{1.5cm}
To determine the critical temperature of the system, we have performed
a standard finite size scaling analysis \cite{Binder}
 based on the distribution of the
 parameter $m_N$  defined as in (\ref{def-order-parameter}). Such an analysis has
 already  proved to be
effective in the study of non-equilibrium steady states \cite{OLE, TP}.
We performed Monte Carlo simulations of the model for system sizes $N = 30$ to
$210$   over $10^9$ sweeps in the steady state for each set
of parameters. 
To determine the critical temperature $T_c$  we measured the ratio 
between two moments of the parameter $m_N$
\be
U_N & = & 1 - \f{\langle  m_N^4\rangle}{3\langle  m_N^2\rangle^2}\quad. \label{cumulant}
\ee
(In fact, since one does not have to distinguish between
positive and negative values of the parameter $m$,
the ratio $\langle m^2 \rangle /\langle m\rangle^2$ could as well be used to determine the critical temperature.)

\begin{figure}[htb]
\begin{center}
\includegraphics[scale=0.3]{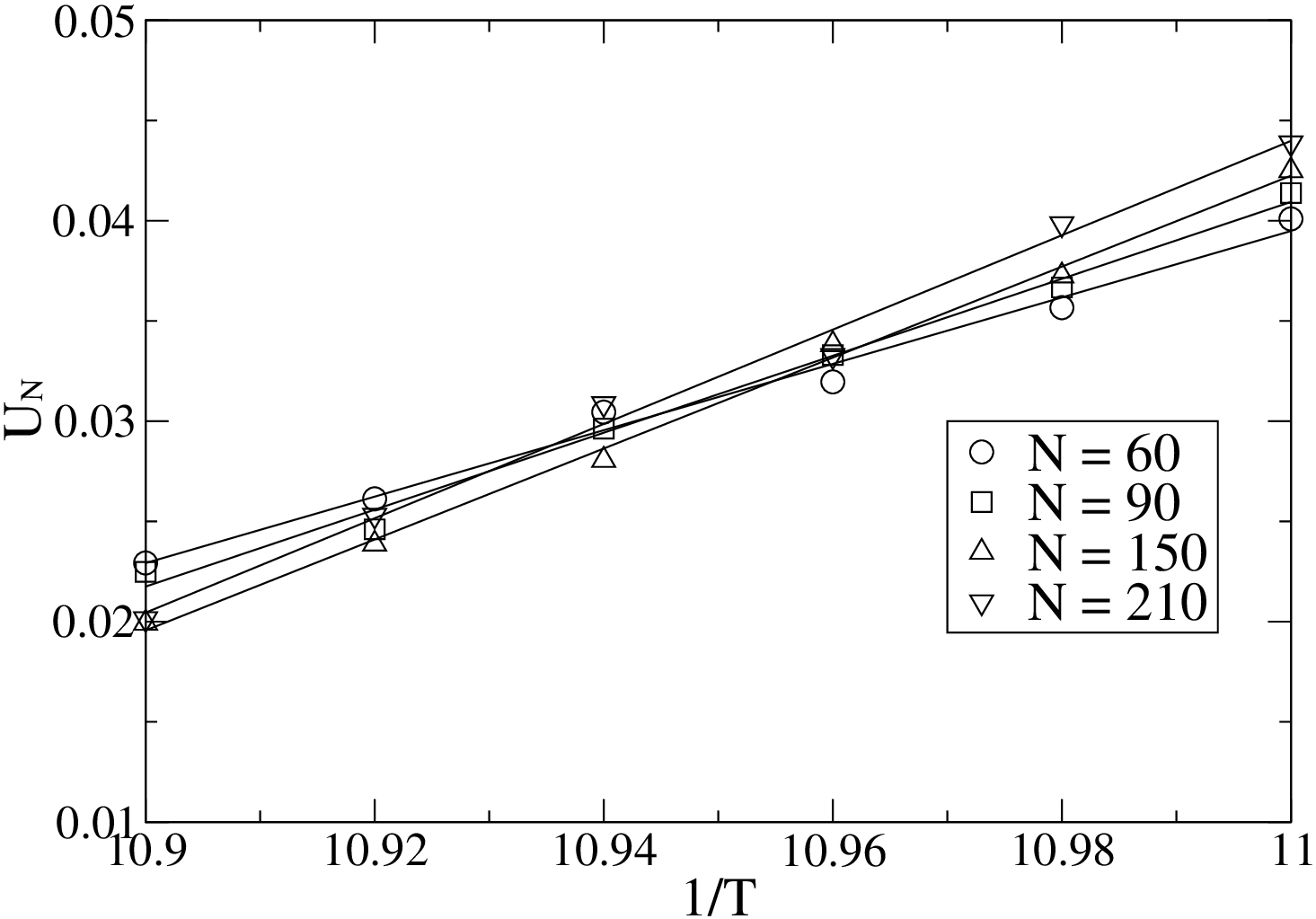}
\caption{Critical temperature $T_c$ as intersection point $U^*$ of the
fourth-order cumulant of the order parameter $U_N$ (the straight lines
are linear fits to the data) }\label{UN-plot}
\end{center}
\end{figure}

At the critical temperature $T_c$, $U_N$ has a universal value $U^*$.
Thus on measuring $U_N$ for various system sizes as a function of $T$,
$U^*$ is the common intersection point of the curves and identifies
$T_c$.  Our results shown in Fig. \ref{UN-plot} indicate $ \beta_c
= 1/T_c \simeq 10.95$.

\vspace*{0.5cm}

\section{Large deviation functional for equal densities and the exact
transition temperature}\label{equal-densities}
In this section we consider only the case 
of equal numbers of each species
\be
N_A = N_B= N_C = {N \over 3}
\ee
where there exists an
energy function given by  (\ref{hamiltonian},\ref{def-energy})
\be 
E_N(\{X_i\}) =\frac{1}{N^2}
\sum_{i=1}^{N-1}\sum_{k=1}^{N-1}
 k \left(B_iC_{i+k}+ C_iA_{i+k}+A_iB_{i+k}\right)
\label{energy}
\ee
with  the occupation variables
$\{X_i\}=
\{A_i, B_i, C_i\}$  defined
as in (\ref{hamiltonian}). 
The steady state probabilities of the system are given by  
(\ref{weights2}).

We pass to the continuum limit where
$\rho_A(x), \rho_B(x), \rho_C(x)$ are the density profiles 
of $A, B, C$ particles
respectively 
at position $x=i/N$.
Then 
the  free energy functional   \cite{DLS1,DLS2} (or large 
deviation functional)  
${\cal F}[\rho_A(x), \rho_B(x), \rho_C(x)]$ which gives
the probability of any  density profile
through
\be
P[\rho_A(x), \rho_B(x), \rho_C(x)] 
 = \exp\{ - N {\cal F}[\rho_A(x), \rho_B(x), \rho_C(x)] \}
\label{pro}
\ee
can be written in terms of the density profiles as
\begin{eqnarray}
\lefteqn{{\cal F}[\rho_A(x), \rho_B(x), \rho_C(x)] = K } \nonumber\\
&+&  \int_{0}^{1} dx \left[
\rho_A(x) \ln \rho_A(x) +\rho_B(x) \ln \rho_B(x)
+\rho_C(x) \ln \rho_C(x) \right]
\nonumber \\
&+&   \beta \int_{0}^{1} dx \int_{0}^{1} dz
\left[ \rho_B(x) \rho_C(x+z) +
\rho_C(x) \rho_A(x+z) + \rho_A(x) \rho_B(x+z)\right] z 
\label{FEF}
\end{eqnarray}
where $\rho_A,\rho_B,\rho_C$ are periodic functions of period $1$ and $K$ is a normalization constant such that the minimum of
${\cal F}$ over all profiles vanishes.
The second term on the right hand side of (\ref{FEF}) 
represents the entropy of the given profiles
while the term  proportional to $\beta$ 
is the continuum form of the energy (\ref{energy}).

It is easy to derive an expression for
the optimal profile for $\rho_A(x)$ by finding the extremum of
(\ref{FEF}) with respect to $\rho_A(x)$ subject to the constraint
that 
$$\int \rho_A(x)dx = 1/3\;.$$
One obtains
\be
\rho_A(x) = \mbox{Constant} \times \exp \left(-\beta
 \int_0^{1} [\rho_B(x+z) +  \rho_C(x-z) ] z dz 
\right)\;.
\ee
which implies that
\be
\frac{d \rho_A(x)}{d x}
= -\beta \rho_A(x) \left( \rho_B(x) - \rho_C(x) \right)\;.
\label{Aeqfirst}
\ee
Similar equations  hold for $\rho_B(x)$
and $\rho_C(x)$ 
and using  $\rho_C(x) = 1- \rho_A(x)-\rho_B(x)$
one obtains  the coupled equations
\begin{eqnarray}
\label{Aeq}
\frac{d \rho_A(x)}{d x}
&=& -\beta \rho_A(x) \left( \rho_A(x) +2 \rho_B(x) -1\right)\\
\label{Beq}
\frac{d \rho_B(x)}{d x}
&=& -\beta \rho_B(x) \left( 1-2\rho_A(x)  - \rho_B(x) \right)\;.
\end{eqnarray}

Clearly one solution of (\ref{Aeq},\ref{Beq})
is $\rho_A(x)=\rho_B(x)=1/3$ which corresponds to the disordered
phase, but this extremum is the minimum only in the disordered phase. To test when an ordered solution emerges
we define the Fourier series of
arbitrary profiles as
\be
 \rho_A(x) &=& 1/3 + \sum_{n=1}^{\infty}\left[
a_n \exp (i 2\pi n x) + a_{-n} \exp (-i2\pi  n x)
\right] \\
 \rho_B(x) &=& 1/3 + \sum_{n=1}^{\infty}
\left[ b_n \exp (i2\pi n x) + b_{-n} \exp (-i2\pi  n x)
\right] \\
 \rho_C(x) &=& 1/3 + \sum_{n=1}^{\infty}
\left[ c_n \exp (i2\pi n x) + c_{-n} \exp (-i2\pi n x) \right]
\label{Fseries}
\ee
We insert these
into (\ref{Aeq},\ref{Beq})  and,
anticipating a continuous phase transition, expand to first order in $a_n,b_n$.
We look for values of $\beta$ for which 
a solution for nonzero
$a_n$, $b_n$, $c_n$ is present and therefore the
uniform solution could be unstable.
One finds that the uniform solution becomes unstable
to  the $n$th mode for $\beta^2/3 =
(2\pi n)^2$  so that the first instability occurs at $\beta_c$ given by
\be
\beta_c =  2\pi \sqrt{3} = 10.882796...\;.
\label{betac}
\ee 
Near $\beta_c$ equations (\ref{Aeq},\ref{Beq}) can be solved
perturbatively in 
\begin{equation}
 \epsilon = \frac{\beta -\beta_c}{\beta_c}\;.
\label{eps}
\end{equation}
 One finds
to leading order 
\be \lefteqn{\rho_A(x)= \rho_B \left(x+{1 \over 3 }
\right) =  \rho_C \left(x+{2 \over 3 } \right) = }\hspace{2cm}\nonumber\\
 &&\frac{1}{3}+ \left(
\epsilon \over 6 \right)^{1/2} 2\cos(2 \pi (x-x_0)) + \left( \epsilon
\over 6 \right) 2\cos(4 \pi (x-x_0)) + O( \epsilon^{3/2 } )
\label{rhoopt}
\ee
where $x_0$ can be arbitrary (as an optimal profile remains optimal when it is translated). 
Also, choosing $x_0=0$, one can show that 
$a_n=a_{-n}$ and $a_{3m} =0\;$ for $|m|= 0,1\ldots$ to all orders in 
$\epsilon$  and 
that to leading order in $\epsilon$
\be
a_1 \sim \left( \frac{\epsilon}{6}\right)^{1/2}\;,\quad
a_2 \sim \left( \frac{\epsilon}{6}\right)\;,\quad
a_4 \sim -\left( \frac{\epsilon}{6}\right)^{2}\;,\quad
a_5 \sim -\left( \frac{\epsilon}{6}\right)^{5/2} \ldots
\label{moderes}
\ee
 The   parameter $m$  defined  as in
(\ref{def-op-limit},\ref{def-order-parameter}) 
 becomes in the continuum limit
\begin{equation}
m =  \int_{0}^1 [ \rho^2_A + \rho^2_B+ \rho^2_C ] dx
 \ -  \ \frac{1}{3}\;.
\end{equation}
and from
(\ref{rhoopt}) one obtains
\be
m \sim  \frac{\beta-\beta_c}{\beta_c}\;.
\label{mfop}
\ee
Thus  the  parameter  $m$ as defined
in (\ref{def-op-limit},\ref{def-order-parameter}) vanishes linearly at
the transition.
An alternative, and probably more standard, choice  for the order
parameter could   be
 the amplitude $a_1$ of the fundamental mode  \cite{KSZ}
 and this would lead to an exponent $1/2$.

We now turn to the calculation of the energy.
For arbitrary density profiles
the  energetic contribution
to the free energy 
(\ref{energy})    may be written
in terms of the Fourier coeffcients
as
\begin{equation}
E = \frac{1}{6} - \sum_{n\neq 0}
\frac{a_n b_{-n} + b_nc_{-n}+c_na_{-n}}{2\pi i n}
\end{equation}
For profiles 
$\rho_B(x) = \rho_A(x-1/3)$,
$\rho_C(x) = \rho_A(x-2/3)$ this  energy becomes
\begin{equation}
E = \frac{1}{6} - 3 \sum_{n=1}^{\infty}
\frac{a_n a_{-n}}{n \pi}\sin (2\pi n/3)\;.
\end{equation}
Thus, using (\ref{moderes}), near $\beta_c$ we have
$E \simeq 1/6 - 3\epsilon/2\beta_c$
and the heat capacity $ -\beta^2 \partial E/\partial \beta$
has a discontinuity of 3/2 at $\beta_c$, consistent with the data of figure \ref{specheat}.

This 
equal density case is similar to some special cases found in  a recent
study of the dynamical winding of random walks\cite{FF} for which
the fact that the dynamics satisfies detailed balance 
allows one to write  equations  for the density profiles of the type
(\ref{Aeq}) and to locate the exact transition point.

\section{Non-equal Densities and mean-field theory}\label{five}
We now turn to the case of non-equal densities of particles. 
A direct consequence of the stochastic dynamical  rules (\ref{update}) is that
\begin{eqnarray}
{d \langle A_i\rangle \over d t} =
q\langle A_{i-1}  B_i\rangle + 
q\langle C_{i}  A_{i+1}\rangle + 
\langle B_{i}  A_{i+1}\rangle + 
\langle A_{i-1}  C_{i}\rangle  \nonumber \\ 
-q\langle A_{i}  B_{i+1}\rangle  
-q\langle C_{i-1}  A_{i}\rangle  
- \langle B_{i-1}  A_{i}\rangle - 
\langle A_{i}  C_{i+1}\rangle  
\label{exact}
\end{eqnarray}
and similar equations hold for ${d \langle B_i\rangle \over d t}$ and ${d
\langle C_i\rangle \over d t}$. 
We do not know how to solve these exact equations, in particular because
they require the knowledge of
two point functions.
One can, however, write down mean-field equations \cite{DDM}, by making an approximation which neglects
correlations (i.e. where one replaces correlation functions such as $\langle A_{i-1}  B_i\rangle$ by $\langle A_{i-1} \rangle \langle  B_i\rangle$)
\begin{eqnarray}
{d \langle A_i\rangle \over d t} =
q\langle A_{i-1} \rangle \langle  B_i\rangle + 
q\langle C_{i}  \rangle \langle A_{i+1}\rangle + 
\langle B_{i} \rangle \langle  A_{i+1}\rangle + 
\langle A_{i-1} \rangle \langle  C_{i}\rangle  \nonumber \\ 
-q\langle A_{i} \rangle \langle  B_{i+1}\rangle  
-q\langle C_{i-1}  \rangle \langle A_{i}\rangle  
- \langle B_{i-1} \rangle \langle  A_{i}\rangle - 
\langle A_{i}  \rangle \langle C_{i+1}\rangle  
\label{mf}
\end{eqnarray}

Assuming that the profiles vary slowly with $i$,
we write $\rho_A(x) = \langle A_i
\rangle$ and
\be
\langle A_{i\pm 1}  \rangle= \rho_A(x) \pm
\frac{1}{N} \frac{\partial \rho_A(x)}{\partial x}
+\frac{1}{2 N^2} \frac{\partial^2 \rho_A(x)}{\partial x^2} +\ldots
\ee
then keeping leading order terms in $1/N$
and defining $\tau = N^2 t$ in (\ref{tdmf}) yields
\begin{eqnarray}
\frac{\partial \rho_A}{\partial \tau} =
\beta \frac{\partial}{\partial x} \left[ \rho_A(\rho_B-\rho_C)\right] +
\frac{\partial^2\rho_A}{\partial x^2} \nonumber\\
\frac{\partial \rho_B}{\partial \tau} =
\beta \frac{\partial}{\partial x} \left[ \rho_B(\rho_C-\rho_A)\right] +
\frac{\partial^2\rho_B}{\partial x^2} \label{tdmf}\\
\frac{\partial \rho_C}{\partial \tau} =
\beta \frac{\partial}{\partial x} \left[ \rho_C(\rho_A-\rho_B)\right] +
\frac{\partial^2\rho_C}{\partial x^2} \nonumber
\end{eqnarray}
One can linearize these equations around constant density profiles
\be
\rho_A(x) = r_A + \Delta \rho_A(x)\;,\quad
\rho_B(x) = r_B + \Delta \rho_B(x)\;,\quad
\rho_C(x) = r_C + \Delta \rho_C(x)
\ee
where $\Delta \rho_A(x),\Delta \rho_B(x), \Delta \rho_C(x)$
represent small departures from constant profiles at densities $r_A=
N_A/N, r_B=N_B/N, r_C=N_C/N$ ($r_A,r_B,r_C$ are the global densities
of the three species and of course they satisfy $r_A+r_B+r_C=1$).
Then one finds that these small departures are damped when $\beta <
\beta_c^{\rm mf}$ given by
\be
\beta_c^{\rm mf} = \frac{ 2\pi}{
\left( 2r_A r_B  +2 r_A r_C + 2 r_B r_C - r_A^2 - r_B^2 - r_C^2  \right)^{1/2}}
= \frac{ 2\pi}{\left[ 1 - 2 (r_A^2 + r_B^2 + r_C^2)\right]^{1/2}}
\label{betamf}
\ee

We see that in the equal density case ($r_A=r_B=r_C= 1/3$), the mean
field value of $\beta_c$ coincides with the exact value
(\ref{betac}). Moreover one finds that the solutions of the exact
equations for the optimal profile (\ref{Aeqfirst}) are steady state
solutions of the mean field equations (\ref{tdmf}) (as they make the
l.h.s. of these equations vanish).  So, at least for the equal time
properties, in this equal density case, $\beta_c$ and the profiles
predicted by the mean field theory are exact.

Near $\beta_c$, one can perturbatively find stationary non-moving
profiles. The first Fourier mode of these profiles
is given by 
\be
 \Delta \rho_A(x) &\simeq& \psi(\epsilon)  \left[\sqrt{r_A}  \  e^{2 i \pi (x-x_0)} +
c.c. \right] 
\label{rhoa}\\[1ex]
\Delta \rho_B(x) &\simeq& \psi(\epsilon)  \left[ {r_C - r_A - r_B - i \sqrt{ 1 - 2
(r_A^2 + r_B^2 + r_C^2) } \over 2 \sqrt{r_A}}   e^{2 i \pi
(x-x_0)} +
c.c. \right] 
\label{rhob}\\[1ex]
\Delta \rho_C(x) &\simeq& \psi(\epsilon)  \left[ {r_B - r_A - r_C + i \sqrt{ 1 - 2
(r_A^2 + r_B^2 + r_C^2) } \over 2 \sqrt{r_A}}   e^{2 i \pi
(x-x_0)} +
c.c. \right] 
\label{rhoc}
\ee
where  $\epsilon$ is defined as in (\ref{eps}) and
\begin{equation}
\psi(\epsilon) =  
 {1 - 2
(r_A^2+r_B^2+r_C^2) \over \sqrt{2 (r_A^2 +
r_B^2 + r_C^2) - 4 (r_A^3 + r_B^3 + r_C^3) }}
\ \epsilon^{1/2}
\label{op}
\end{equation}

Analyzing the whole phase diagram predicted by the mean-field
equations (\ref{tdmf}) is not an easy task. Apart from the constant
profile solutions, which become unstable for $\beta > \beta_c^{\rm
mf}$, other (static or moving) solutions might exist in some regions
of the phase diagram and a full description of the phase diagram would
require the knowledge of all the solutions of the mean field equations
and of their stabilities.

Equation (\ref{betamf}) implies that for
$r_A^2 + r_B^2 + r_C^2 > 1/2$ there is no second order phase
transition. However, looking at (\ref{op}) it is clear that the second order
transition from the flat profile solution to the  solution
(\ref{rhoa}-\ref{op}) should already  become first order when
$r_A^2 + r_B^2 + r_C^2  < 2 (r_A^3 + r_B^3 + r_C^3)$.
This is similar to what occurs in a mean-field study of another
lattice gas\cite{VSZ}.

At the moment we cannot tell  whether the  predictions of the mean field
theory (\ref{tdmf}) such as 
(\ref{betamf}) remain exact in the  
case of unequal global densities.
To try to shed some light on
this question,  
we have  calculated the large deviation functional to order $\beta^2$
in a small $\beta$ expansion. The details given in the appendix
show that, in the unequal density case, the large deviation functional
(\ref{pro},\ref{FEF}) becomes 
\begin{eqnarray}
\lefteqn{{\cal F}[\rho_A(x), \rho_B(x), \rho_C(x)] = K 
+  \int_{0}^{1} dx \left[
\rho_A(x) \ln \rho_A(x) +\rho_B(x) \ln \rho_B(x)
+\rho_C(x) \ln \rho_C(x) \right] }\hspace{0.5cm} \nonumber\\
&+ &   \beta \int_{0}^{1} dx \int_{0}^{1}  dz \ z
\left[ \rho_B(x) \rho_C(x+z) +
\rho_C(x) \rho_A(x+z) + \rho_A(x) \rho_B(x+z)\right]  \nonumber \\
&-& {3 \over 4}  \beta^2 \int_{0}^{1} dx \int_0^1 dz \;  z (1-z) \left[ 
 r_A (1 - 3 r_A) \rho_B(x) \rho_C(x+z) + r_B (1 - 3 r_B)\rho_C(x) \rho_A(x+z) \right.\nonumber \\
& & \qquad\qquad\qquad\qquad\left.+ r_C(1-3 r_C)\rho_A(x) \rho_B(x+z)
\right]  
+ O( \beta^3)
\label{FEFnew}
\end{eqnarray}
We see that for the unequal density case, the large deviation functional
is modified and terms of order $\beta^2$ appear which were not present in
(\ref{FEF}). This $\beta^2$ term is  the first of a whole series in
$\beta$. Without knowing these higher order terms, it is neither possible to
predict the exact value of $\beta_c$ nor  how
 the equations (\ref{Aeqfirst},\ref{Aeq},\ref{Beq}) for the
most likely profiles would be modified. 
 One cannot exclude the possibility that the most
likely profile in the ordered phase corresponds to  moving domains.

\section{Conclusion}
In this  work we have investigated a locally driven system of three
species on a ring which exhibits anomalous coarsening into a strongly
phase separated steady state. In the case of equal numbers of the
particles of each species the steady state obeys detailed balance with
respect to a long-range superextensive energy function, despite the
strictly nearest neighbour dynamics.  In the
weak asymmetry regime where $q\to 1$  as in (\ref{scale1}) one  recovers
an extensive
energy.  In this  regime we found a second-order phase
transition.

For the case of equal particle densities
we have derived the large deviation (or free energy) functional
(\ref{pro},\ref{FEF}) for the profiles.  
As in other examples of non-equilibrium systems studied recently \cite{DLS1,DLS2},
 this functional is non-local
 allowing phase
transitions in one dimension.
 By minimising this  free energy functional
we have  obtained the equations (\ref{Aeqfirst})  satisfied by the optimal density profiles and analysed the phase
transition in the equal density case, in particular we found $\beta_c= 2
\pi \sqrt{3}$.

For the general case of arbitrary particle densities we used a
mean-field theory.  In the case of equal particle densities it turns
out that the mean-field solution yields the same profiles as the exact
free energy optimisation just discussed.  An open question  remains
as to what extent mean field theory is valid when the densities are
unequal.

The mechanism for the phase separation that we have studied can be
understood through the stability of the Fourier modes of the particle
densities. In the high temperature phase, the system is disordered and
the constant density profiles are stable. $T_c$ denotes the temperature at
which the lowest Fourier mode becomes unstable. 
As $T$ decreases, more and more modes become unstable and
the depth of the quench from the disordered high temperature phase into the  low
temperature phase determines the number of
unstable modes which can grow from the constant profile solution.
  However, in the steady state 
 one expects only   three pure domains. 
How the non-linear evolution  of the excited modes combined
with the effect of the noisy dynamics  
determine the anomalous coarsening \cite{EKKM} towards the three pure
domains is  another
interesting open question.

\section{APPENDIX: Expanding the steady state in powers of the bias}
In this appendix, we justify the expression (\ref{FEFnew}) of the large
deviation functional.
Let us consider a finite system of $N$ lattice sites and write the
asymmetry $q$  as
$$q= e^{-\phi} \ . $$
We are going to show in this appendix  that the unnormalized weight  $P({\cal C})$ of a configuration  ${\cal
C}$ in the steady state
can be written  to order $\phi^2$ as
\be
P({\cal C }) = \exp[ \phi R_1 ({\cal C })+ \phi^2 R_2({\cal C
})] \label{PC} 
\ee
where 
\begin{equation}
R_1({\cal C }) = -  { 1 \over N} \sum_{i=1}^N \sum_{d=1}^{N-1} d \ 
( B_i C_{i+d} +  C_i A_{i+d} +  A_i B_{i+d})
\label{R1}
\end{equation}
and
\begin{eqnarray}
\lefteqn{R_2({\cal C }) =   {3 \over  4 N^2 (N-1)} \sum_{i=1}^N
\sum_{d=1}^{N-1} d (N-d)}\hspace{1cm}\nonumber\\      
&&\times \left[ N_A( N- 3 N_A) B_i C_{i+d} 
 + N_B( N- 3 N_B) C_i A_{i+d} + N_C( N- 3 N_C)  A_i B_{i+d}\right]
\label{R2}
\end{eqnarray}
In the weak asymmetry regime (\ref{scale1}) this leads to the formula
(\ref{FEFnew}) in the large $N$ limit.

Let us try to compare two configurations ${\cal C }$ and ${\cal C'}$
of the form
\begin{eqnarray}
 {\cal C  }= X_1 X_2 ...X_{i-1} A_i B_{i+1} X_{i+2} .... X_N 
\label{config} \\
 {\cal C' }= X_1 X_2 ...X_{i-1} B_i A_{i+1} X_{i+2} .... X_N \nonumber
\end{eqnarray}
which differ only by an exchange of a $A$ and a $B$ particle
between sites $i$ and $i+1$ (the notation in (\ref{config}) means that
site $i$ is occupied by a $A$ particle, site $i+1$ by a $B$ particle and 
all the other sites are occupied by arbitrary particles).

If (\ref{R1},\ref{R2}) were true,  we would have
\be
R_1({\cal C' })= R_1({\cal C }) - 3 {N_C \over N} 
\label{R1a}
\ee
\begin{eqnarray}
R_2({\cal C' })= R_2({\cal C })   +  {3 \over 4 N^2 (N-1)} \left[
 N_A (N-  3 N_A ) \sum_{d=2}^{N-1} C_{i+d} (N-2d+1) \right. \nonumber \\
 - N_B (N- 3 N_B ) \sum_{d=2}^{N-1} C_{i+d} (N-2d+1) 
+ N_C (N- 3 N_C ) \sum_{d=2}^{N-1} A_{i+d} (N-2d+1) 
\nonumber \\
 \left.- N_C (N- 3 N_C ) \sum_{d=2}^{N-1} B_{i+d} (N-2d+1)  \right]
\label{R2a}
\end{eqnarray}
 Using the fact that
$$C_{i+d} = 1 - B_{i+d} - A_{i+d} $$
this can be rewritten as
\begin{eqnarray}
R_2({\cal C' })= R_2({\cal C })   + 
 Q_A \sum_{d=2}^{N-1} A_{i+d} (N-2d+1)
 -Q_B \sum_{d=2}^{N-1} B_{i+d} (N-2d+1)  
\label{diff}
\end{eqnarray}
where 
\begin{eqnarray}
Q_A &=& {3 \over 4} \  { N_C( N - 3 N_C) + N_B ( N - 3 N_B) - N_A ( N - 3
N_A) \over N^2 (N-1)}  \\
Q_B &=& {3 \over 4} \ { N_A ( N - 3 N_A) + N_C ( N - 3 N_C) -N_B( N - 3
N_B) \over N^2 (N-1) }
\end{eqnarray}
and $Q_C$ is similarly defined.
Thus exchanging the particles
at an $AB$ interface produces in (\ref{diff}) a difference between a term
involving only A particles and a term
involving only B particles.

Let us consider now a cluster of $m$ sites occupied by $A$ particles
so that ${\cal C }$ has the form
$$ {\cal C  }= X_1 X_2\dots X_{i} A_{i+1} \ldots  A_{i+m} X_{i+m+1}\ldots X_N$$
where  sites $i$ and $i+m+1$ are occupied by particles $B$ or $C$,
so that
$$X_i = L'\qquad\mbox{and}\qquad
X_{i+m+1}=L''$$ 
where $L'$ is a $B$ or a $C$ particle and so is $L''$.

There are two configurations ${\cal C' }$ and ${\cal C'' }$ which can be reached from
${\cal C  }$ by single moves at the two boundaries of this cluster.
A rather simple calculation using (\ref{R2a}) shows that
\begin{eqnarray}
\lefteqn{R_2({\cal C' })+ R_2({\cal C'' })- 2 R_2({\cal C })=
Q_A [ 2 m N_A - 2 - 2(m-1) N]}\hspace{2cm}
\nonumber \\
&& +Q_{L''} \sum_{d=2}^{N-1} {L''}_{i+m+d} (N-2d+1) 
 -Q_{L'} \sum_{d=2}^{N-1} {L'}_{i+d} (N-2d+1)  \;.
\label{R2b}
\end{eqnarray}
Then summing over all clusters yields
\begin{eqnarray}
\lefteqn{\sum_{{\cal C'}}\left[ R_2( {\cal C'}) - R_2({\cal C})\right]=  
 2 Q_A [  N_A^2  - N N_A + (N-1)(   N_{AB} + N_{AC} )]}\hspace{0.5cm}
 \\
&&+  2 Q_B [  N_B^2 -  N N_B + (N-1)( N_{BC} + N_{BA} )]
+  2 Q_C [  N_C^2 -  N N_C + (N-1)(  N_{CA} + N_{CB} )]\nonumber
\end{eqnarray}
where the sum is over all the configurations ${\cal C'}$ which
can be reached from a given  ${\cal C}$ by a single exchange
and $N_{AB}, N_{BA}, N_{BC},...$ are the numbers of neighboring pairs
$AB,BA,BC...$ along the chain.

So far (\ref{R1a},\ref{R2b}) have been derived assuming that
(\ref{R1},\ref{R2}) are true. To prove that (\ref{PC}) does give the correct steady
state weights, one
needs to check stationarity, i.e.  that
\begin{eqnarray}
\lefteqn{\left(1 - \phi     + {\phi^2 \over 2 } \right) (N_{AB} + N_{BC} + N_{CA} )
+ N_{BA} + N_{CB} + N_{AC}  =}\hspace{1.0cm}
 \nonumber \\ 
&&\left(1 - \phi + {\phi^2 \over 2 } \right)
 \left(N_{BA} e^{3 \phi N_C \over N }
 + N_{CB}e^{3 \phi N_A \over N } + N_{AC}e^{3 \phi N_B \over N } \right) 
 \nonumber \\ 
&+&\left(N_{AB} e^{-3 \phi N_C \over N }
 + N_{BC}e^{-3 \phi N_A \over N } + N_{CA}e^{-3 \phi N_B \over N } \right) 
+ \phi^2 \sum_{{\cal C'}}  \left[ R_2( {\cal C'}) - R_2({\cal C})  \right]
\end{eqnarray}
This can be checked  to order $\phi^2$,  using simply the fact that
for any configuration
$$N_{AB} -N_{BA} = N_{BC} - N_{CB} = N_{CA} - N_{AC}\;.$$

\begin{acknowledgments}
MC would like to thank the Gottlieb Daimler - und Karl Benz - Stiftung, DAAD 
as well as EPSRC for financial support.
\end{acknowledgments}

\bibliography{Amine}

\end{document}